\documentstyle{mn}

\def \mnras {MNRAS}
\def \apj {ApJ}
\def \aap {A\&A}

\begin{document}
\title [Lagrangian Approximations in Voids]
{Behaviour of Lagrangian Approximations in Spherical Voids}

\author [V. Sahni and S.F. Shandarin]
{Varun Sahni$^1$ and Sergei Shandarin$^2$\\
$^1$Inter-University Centre for Astronomy and Astrophysics
Post Bag 4, Ganeshkhind, Pune 411 007, India\\
$^2$Department of Physics and Astronomy, University of Kansas, U.S.A.}
\maketitle
\begin {abstract}
We study the behaviour of spherical Voids in Lagrangian perturbation theories
L(n), of which the Zel'dovich approximation is the lowest order solution L(1).
We find that at early times higher order L(n) give an increasingly
accurate picture of Void expansion.
However at late times particle trajectories
in L(2) begin to turnaround and converge leading to the
{\em contraction} of a Void, a sign of pathological behaviour.
By contrast particle trajectories in L(3) are well behaved
and this approximation gives results in excellent agreement
with the exact top-hat solution as long as the Void is not too underdense.
For very underdense Voids, L(3) evacuates the Void much too rapidly
leading us to conclude that the Zel'dovich approximation L(1),
remains the best approximation to apply to the late time study of Voids.
The behavior of high order approximations in spherical voids is
typical for asymptotic series and may be generic for Lagrangian
perturbation theory.
\end {abstract}


\section {Introduction}
It is widely felt that gravitationally bound systems such as galaxies and
groups and clusters of galaxies arose due to the gravitational amplification
of small density fluctuations already existing at the time of last scattering
(and reflected in the angular fluctuations in the CMBR measured by the COBE
satellite).

The growth of density perturbations in a homogeneous and isotropic Universe
can be studied in essentially two distinct ways. In the {\it Eulerian} approach
(pioneered by Jeans and Lifshitz), solutions to the Euler-Poisson system
are sought
by means of a dimensionless density perturbation
$\delta({\bf x}, t)=[\rho({\bf x}, t)-\rho_0]/\rho_0$, which is expanded in
a perturbation series having the form
\begin{equation}
\delta({\bf x},t) =  \sum_{n=1}^{\infty} \delta^{(n)} ({\bf x},t)
 = \sum_{i=1}^{\infty} D_{+}^n (t) \Delta^{(n)}({\bf x})
\label{eq:euler}
\end{equation}
in a spatially flat matter dominated Universe
(similar expansions are made for the velocity field and gravitational
potential).
$\delta^{(1)}$ corresponds to the linear theory limit for which
$D_{+}(t) \propto a(t) \propto t^{2/3}$. Eulerian perturbation theory can be
successfully applied as long as the
series (\ref{eq:euler}) converges {\it i.e.} $|\delta| << 1$.

An alternate viewpoint to Eulerian theory was proposed by Zel'dovich who
suggested perturbing particle trajectories as a means to understand
gravitational
instability. In a general {\em Lagrangian} framework
\begin{equation}
{\bf x} = {\bf q} + {\bf \Psi}(t, \bf q)
\label{eq:la1}
\end{equation}
where ${\bf x}$ and ${\bf q}$ are the Eulerian and Lagrangian coordinates of
a particle/fluid element. The displacement vector
${\bf \Psi}$ is related to the initial velocity
field and its derivatives, in Zeldovich's approximation
${\bf \Psi} = {\bf \Psi}^{(1)} = D_+(t)\psi_i^{(1)}({\bf q}),~$
$\psi_i^{(1)}$ being the initial velocity field.
In recent years attempts have been made to consider higher order corrections
to the Zel'dovich approximation by treating ${\bf \Psi}$ as an expansion
(Moutarde et al. 1991, Bouchet et al. 1992, Buchert 1993, Catelan 1995,
Bouchet et al. 1995)
\begin{equation}
{\bf \Psi} = \sum_{i=1}^{n}{\bf \Psi}^{(n)}
\label{eq:la2}
\end{equation}
Expansions of the form (\ref{eq:la1}), (\ref{eq:la2}) are
frequently referred to
as {\it n-th order} Lagrangian perturbation series L(n),
the first order series L(1) being the Zel'dovich approximation.
One might expect (in analogy with Eulerian perturbation theory -- E(n)) L(n) to
get more accurate with increasing `n' as higher order terms are included in
$\Psi$ (Coles \& Sahni 1995).
This is indeed the case during the weakly non-linear regime as
demonstrated by Munshi, Sahni \& Starobinsky (1994)
who showed that results derived from
L(n) matched those derived from E(n) as long as the system is weakly nonlinear.
An important issue which has so far remained largely unaddressed concerns the
domain of convergence of the L(n) series (\ref{eq:la1}) \& (\ref{eq:la2}).
Although several recent studies have demonstrated the increasing accuracy of
higher order
L(n), most treatments (with the exception of Bouchet et al. 1995)
simulated overdense regions and left unanswered
the related (and important) issue of the accuracy of L(n) in underdense
regions (Voids). We shall address this issue in this letter by
studying the behavior of L(n) in spherical underdense regions
complementing and extending the analysis of Munshi et al. (1994)
for overdense regions.
(We leave a more general analysis treating generic initial conditions
to a later work (Buchert, Sahni \& Shandarin 1995)).

\section {Lagrangian Perturbation Theory in Voids}
In a spatially flat, matter dominated Universe $\Psi_i^{(n)}$ factorise:
\begin{equation}
\Psi_i^{(n)}=D_+^n(t)\psi_i^{(n)}({\bf q}),~~~~
D_+(t)= {3t^2\over 2a^2}\propto a(t)\propto t^{2/3}
\label{eq:la3}
\end{equation}
Furthermore higher orders in ${\bf \Psi}^{(n)}$ ($n>1$) can be constructed
from lower orders by means of an iterative
procedure (Moutarde et al. 1991, Buchert 1992, Lachieze-Rey 1993), so that
$\psi_i^{(1)}= -\partial \phi_0({\bf q})/ \partial q_i$, for L(1)
(the Zel'dovich approximation),
for L(2) (equivalently the {\em post-Zel'dovich approximation})
(Bouchet et al. 1992, Buchert 1992)
\begin{equation}
\psi_{i,i}^{(2)} = -{3\over 14}\left((\psi_{i,i}^{(1)})^2
- \psi_{i,j}^{(1)}\psi_{j,i}^{(1)}\right)
\label{eq:pza1}
\end{equation}
\begin{equation}
\psi_{i,j}^{(2)} = \psi_{j,i}^{(2)}
\label{eq:pza2}
\end{equation}
For L(3):
\begin{eqnarray}
&\psi_{i,i}^{(3)} =  -{5\over 9}(\psi_{i,i}^{(2)} \psi_{j,j}^{(1)}
- \psi_{i,j}^{(2)} \psi_{j,i}^{(1)}) - {1\over 3}det\lbrack \psi_{i,j}^{(1)}
\rbrack, \nonumber\\
&\psi_{i,j}^{(3)} - \psi_{j,i}^{(3)} = {1\over 3}(\psi_{i,k}^{(2)}
\psi_{k,j}^{(1)} - \psi_{j,k}^{(2)} \psi_{k,i}^{(1)})
\label{eq:pza3}
\end{eqnarray}
Here coma implies partial derivative with respect to ${\bf q}$, summation
over repeated indices is assumed (cf. Juszkiewicz et al. 1993, Bernardeau
1993).

Let us consider the case of a spherical top-hat underdensity for which
$\delta^{(1)}({\bf x},t) = - a(t)$, the initial gravitational potential can
be obtained from the Poisson equation and has the time independent form
(Munshi et al. 1994)
$\phi_0= a^2r_0^2\delta/9t^2 = - a^3r_0^2/9t^2$.

The exact solution for the expansion of a top-hat Void can be parametrised as
\begin{eqnarray}
R(\theta) & = & (3/10)(\cosh\theta-1)\nonumber\\
a(\theta) & = & (3/5)\Big [(3/4)(\sinh\theta-\theta)\Big ]^{2/3}
\label{eq:th1}
\end{eqnarray}
Here $R = a(t)r/r_0$ is the physical (and $r$ the comoving) particle
trajectory.
Knowing $r(a)$
the equation of mass conservation allows one
to determine the density contrast
\begin{equation}
\delta=(r_0/r)^3 - 1
\label{eq:cons}
\end{equation}
which for negative density fluctuations has the form
\begin{equation}
\delta_{TH}(\theta) = {9\over 2}{(\theta - \sinh\theta)^2\over (\cosh\theta -
1)^3}
- 1
\label{eq:th2}
\end{equation}
The related dimensionless peculiar velocity
$(1/ aH)~ div_{\bf x} {\bf u} =\theta$ (${\bf u} = a\dot{\bf x}$,
${\bf x}$ is the comoving
coordinate) may be determined from the Eulerian mass conservation equation
(Munshi et al. 1994)
\begin{equation}
a {d\over da} \delta  + (1 + \delta)\theta = 0
\label{eq:theta}
\end{equation}

Particle trajectories for spherical top-hat expansion are easy to derive in
L(n), substituting $\phi_0= - a^3r_0^2/9t^2$ in (\ref{eq:pza2}) \&
(\ref{eq:pza3})
we get
\begin{eqnarray}
r_1(a) = r_0(1 + {a\over 3}) \nonumber\\
r_2(a) = r_0\left(1 + {a\over 3} - {a^2 \over 21}\right)\nonumber\\
r_3(a) = r_0\left(1 + {a\over 3} - {a^2 \over 21} + {23a^3\over 1701}\right)
\label{eq:traj}
\end{eqnarray}
$r_n$ is the particle trajectory in L(n).
{}From (\ref{eq:traj}) and the conservation condition (\ref{eq:cons})
we obtain expressions for the density in a top-hat void
\begin{eqnarray}
\delta_1 = \left(1 + {a \over 3}\right)^{-3} - 1\nonumber\\
\delta_2 = \left (1 + {a \over 3}- {a^2 \over 21} \right)^{-3} -1\nonumber\\
\delta_3 = \left (1 + {a \over 3}- {a^2 \over 21}
+ {23 a^3\over 1701}\right)^{-3} -1 \nonumber\\
\label{eq:dens}
\end{eqnarray}
The related expressions for the velocity field are determined from
(\ref{eq:theta})
\begin{eqnarray}
\theta_1 = a \left(1 + {a \over 3}\right)^{-1} \nonumber\\
\theta_2 = (a - {2\over 7}a^2)/(1 + {a \over 3} -
{a^2 \over 21}) \nonumber\\
\theta_3 = (a - {2\over 7}a^2 + {23\over 189}a^3)/
(1 + {a \over 3} - {a^2 \over 21} + {23\over 1701}a^3)
\label{eq:velo}
\end{eqnarray}

(The expressions for a top-hat overdensity can be obtained
from (\ref{eq:traj}) --
(\ref{eq:velo}) by the transformation $a \rightarrow -a$.)

Our results for the evolution of overdense and underdense regions
are summarised in figures (\ref{fig:den}) \& (\ref{fig:vel})
for top-hat density and velocity fields respectively.
Looking at the right-hand side of figure (\ref{fig:den}) (and the
left hand side of figure (\ref{fig:vel})), we see that higher
order L(n) outperform lower orders in matching the results of exact top-hat
{\em collapse}.
The remaining two panels -- showing top-hat Void {\em expansion},
however give a strikingly different picture. We find that although
L(2) and L(3) do paint an accurate picture at early times,
their accuracy declines once the top-hat density falls below
$\delta_{TH} \approx - 0.7$.
In fact, for L(2) particle
trajectories begin to converge (rather than diverge) at late times,
leading to the
`collapse' of a top-hat Void, a rather pathalogical result !
(also see Bouchet et al. 1995).
Third order fares considerably better than second order
since particle trajectories always diverge leading to Void expansion.
However, as we see from figures (\ref{fig:den}) and (\ref{fig:vel}),
the third order Void expands too rapidly at late times,
causing the asymptotic limit $\delta \rightarrow - 1$ to be reached
much too early. (Interestingly $\delta_2(a) > \delta_{TH}(a)$,
whereas $\delta_1(a), \delta_3(a) < \delta_{TH}(a)$.)
(In simulations with random initial conditions one might expect
shell crossing to occur before the density contrast in Voids has dropped
to $\delta_{TH} \approx -0.7$, as a result the pathological behaviour of
L(2)
may not be discernable in simulations which evolve particles till the
epoch of shell crossing but no further (Buchert, Melott \& Weiss 1994).)

Although the above discussion has been limited to homogeneous Voids, our
main conclusions appear to be more general. To demonstrate this we consider
the case of a spherical density perturbation with an {\em arbitrary} initial
potential $\phi_0(r_0)$.
Expressions for L(1), L(2) and L(3) turn out
to be remarkably simple:
$\vec{\psi}_i^{(1)} = - (\phi_0 '/r_0)~\vec{r}_0$,
$~~~\vec{\psi}_i^{(2)} = -3/7 (\phi_0 '/r_0)^2~\vec{r}_0,$
$~~~\vec{\psi}_i^{(3)} = -23/63 (\phi_0 '/r_0)^3~\vec{r}_0,$
where $\phi_0 ' = \partial \phi_0/ \partial r_0$.
As a result
\begin{eqnarray}
\vec{r_1} = \vec{r_0}\Big [1 - (D_+\phi_0 '/r_0)\Big]\nonumber\\
\vec{r_2} = \vec{r_0}\Big [1 - (D_+\phi_0 '/r_0) - {3\over 7}(D_+\phi_0
'/r_0)^2\Big]\nonumber\\
\vec{r_3} = \vec{r_0}\Big [1 - (D_+\phi_0 '/r_0) - {3\over 7}(D_+\phi_0
'/r_0)^2
- {23\over 63}(D_+\phi_0 '/r_0)^3 \Big]
\end{eqnarray}
(Related expressions for the density and velocity fields can be easily
determined from
(\ref{eq:cons}) and (\ref{eq:theta}).)
Specialising to power
law potentials $\phi_0 = - Ar_0^n$, $A = a^3/9t^2 = constant$, we get
\begin{equation}
\vec{r_2} = \vec{r_0}\Big [1 + nD_+Ar_0^{n-2} - {3\over
7}(nD_+Ar_0^{n-2})^2\Big ]
\label{eq:power}
\end{equation}
($D_+A = a/6$).
We find that for L(2) the displacement vector's
$\vec{\psi}_i^{(2)}$ and $\vec{\psi}_i^{(1)}$ have opposite directions leading
to expansion at early times being replaced by collapse at later epochs.
The form of (\ref{eq:pza1}), (\ref{eq:pza2}) suggests that
this behavior may be generic. Indeed
the transformation $\phi_0 \rightarrow - \phi_0$ (transforming a
concave potential into a convex one) results in
$\vec{\psi}_i^{(1)} \rightarrow -\vec{\psi}_i^{(1)}$,
$\vec{\psi}_i^{(2)} \rightarrow \vec{\psi}_i^{(2)}$, the fact that the
displacement field $\vec{\psi}_i^{(2)}$ remains inwardly directed regardless
of the sign of the potential, results in particles flowing inwards at
late times leading to the eventual `collapse of Voids' in L(2).
Thus we find that although L(2) and L(3) start out being more accurate
than L(1) at the commencement of expansion, their accuracy
declines with time, and somewhat surprisingly, L(1)
provides a more accurate picture of Void expansion than either L(2) or L(3)
at very late times (see figures (\ref{fig:den}) \& (\ref{fig:vel})).

It would be interesting to determine whether the pathalogical behaviour of L(2)
afflicts
all even orders of the Lagrange perturbation series
or whether it is restricted to second order alone.
Although a completely general treatment of this kind lies beyond the scope
of the present letter, it is possible to assess what happens
{\it at any arbitrary
order} for the case of a top-hat Void. This case is exactly solvable
since an expression for the particle
trajectory in L(n) can be found by expanding the
exact top-hat solution (\ref{eq:th1}) in powers of $a$
(Munshi et al. 1994), leading to
\begin{equation}
\vec{r_n} = \vec{r_0}\Big [1 - \sum_{i=1}^n (-1)^i\alpha_ia^i\Big ]
\label{eq:4lpt}
\end{equation}
($\alpha > 0$; for $n = 5$,
$\alpha_1 = 1/3$, $\alpha_2 = 1/21$, $\alpha_3 = 23/1701$,
$\alpha_4 = 1894/392931$, $\alpha_5 = 3293/1702701$.)
The form of (\ref{eq:4lpt}) indicates that all even order terms
will have negative signs leading to the eventual contraction of top-hat Voids
in L(n=even). Odd orders are better than even orders but overestimate
Void expansion at late times giving rise to very empty Voids as demonstrated
in figures (\ref{fig:den}) \& (\ref{fig:vel}) for L(1) \& L(3).

\section {Conclusions}
We have discussed the evolution of spherical Voids in the framework of
Lagrangian Perturbation series L(n) of which the Zel'dovich approximation
is the first order solution.
We find that L(n) with n-even overestimate the density in Voids, whereas
L(n) with n-odd underestimate it.
On the whole we find that L(n=odd) provide better descriptors of
Void expansion than L(n=even). The Zel'dovich
solution L(1) outperforms L(3), L(5),... at late times
which is typical for asymptotic or semiconvergent series. We have not
proved it but speculate that this may be a generic property of the
Lagrangian and probably Eulerian perturbative theories of gravitational
instability.

It should be mentioned that the relative accuracy of
L(1) in spherical top-hat Voids has earlier been tested against
the accuracy of
several Eulerian approximations (EA)
by Sahni \& Coles (1995) and Bouchet et al. (1995),
who demonstrated that L(1) performed significantly better than EA at all times.
The results of this letter strengthen that conclusion and show
that the Zel'dovich approximation
is the best non-linear approximation
to apply to the late time study of spherical Voids
(also see Sahni, Sathyaprakash \& Shandarin 1994).

\bigskip
\noindent {\bf Acknowledgments:}
The authors acknowledge useful discussions with Thomas Buchert, Dipak Munshi
and Alexei Starobinsky.
Acknowledgements are also due to the Smithsonian Institution, Washington,
USA, for International travel assistance under the ongoing Indo-US
exchange program at IUCAA, Pune, India. S. Shandarin acknowledges the
support from NSF Grant AST-9021414, NASA Grant NAGW-3832 and
a University of Kansas GRF 95 Grant.

\vfill\eject

\begin{figure}
\caption{The density contrast $\delta_{APP}$ in Lagrangian perturbation series
L(n) is shown plotted against
the exact top-hat solution $\delta_{EX}$ for underdense regions (lower left)
and overdense regions (upper right).
We find that whereas the accuracy of L(n) increases with `n' when describing
the behaviour of overdense regions, L(n) with $n > 1$, do not fare as
well when applied to underdense regions or `Voids'.
For Voids we notice that although
L(n) with $n = 2, 3$ are initially more accurate than L(1)
(Zel'dovich approximation), their accuracy worsens  with
time. We also find that L(2) shows pathalogical behaviour
at late times when $\delta_{EX} < -0.7$.
(Note: Different scaling has been used to label
the negative and positive axes.)
}
\label{fig:den}
\end{figure}

\begin{figure}
\caption{The dimensionless velocity field $\theta_{APP}$
in Lagrangian perturbation
series L(n)
is shown plotted against the exact solution $\theta_{EX}$,
for overdense regions (lower left)
and underdense regions (upper right). We find that L(n), $n = 2, 3$
give better results than L(1) for overdense but not for underdense
regions. (Note: Different scaling has been used to label
the negative and positive axes.)
}
\label{fig:vel}
\end{figure}

\end{document}